\documentclass[lettersize,journal]{IEEEtran}
\usepackage{amsmath,amsfonts}
\usepackage{algorithmic}
\usepackage{algorithm}
\usepackage{array}
\usepackage[caption=false,font=normalsize,labelfont=sf,textfont=sf]{subfig}
\usepackage{textcomp}
\usepackage{stfloats}
\usepackage{url}
\usepackage{verbatim}
\usepackage{graphicx}
\usepackage{cite}
\usepackage{multirow}
\usepackage{booktabs} % 美观的表格线条
\usepackage{caption} % 支持表格标题设置
\usepackage{array} % 调整列宽
\usepackage[numbers]{natbib} % 使用 natbib 替代 cite
\usepackage[colorlinks=true, citecolor=blue, linkcolor=blue, urlcolor=blue]{hyperref}

\begin{document}
\title{TrustDataFilter:Leveraging Trusted Knowledge Base Data for More Effective Filtering of Unknown Information}
\author{Jinghong Zhang,
        Yidong Cui,
        Weilin Wang,
        Xianyou Cheng
        % <-this % stops a space        % <-this % stops a space
\thanks{Jinghong Zhang, Yidong Cui, and Weiling Wang are with the Beijing University of Posts and Telecommunications, Beijing, China,jsj\_zjh@bupt.edu.cn
Xianyou Cheng is with the State Key Laboratory of Protection for Civilian, Research Institute of Chemical Defense, Beijing, China.(Corresponding author),chengxy2000@163.com}}% <-this % stops a space

\markboth{Journal of \LaTeX\ Class Files,~Vol.~18, No.~9, September~2020}%
{How to Use the IEEEtran \LaTeX \ Templates}

\maketitle

\begin{abstract}
With the advancement of technology and changes in the market, the demand for the construction of domain-specific knowledge bases has been increasing, either to improve model performance or to promote enterprise innovation and competitiveness\cite{ref1}. The construction of domain-specific knowledge bases typically relies on web crawlers or existing industry databases\cite{ref2}, leading to problems with accuracy and consistency of the data. To address these challenges, we considered the characteristics of domain data, where internal knowledge is interconnected, and proposed the Self-Natural Language Inference Data Filtering (self-nli-TDF) framework. This framework compares trusted filtered knowledge with the data to be filtered, deducing the reasoning relationship between them, thus improving filtering performance\cite{ref3}. The framework uses plug-and-play large language models for trustworthiness assessment and employs the RoBERTa-MNLI\cite{ref4} model from the NLI domain for reasoning. We constructed three datasets in the domains of biology, radiation, and science, and conducted experiments using RoBERTa\cite{ref4}, GPT3.5\cite{ref6}, and the local Qwen2 model\cite{ref5}. The experimental results show that this framework improves filter quality, producing more consistent and reliable filtering results.
\end{abstract}


\section{Introduction}
\IEEEPARstart{W}{ith} technological advancements and market shifts, the demand for building domain-specific knowledge bases is increasing to enhance model performance, drive innovation, and strengthen competitiveness. Expanding knowledge bases into specific domains can significantly improve the accuracy in understanding and responding to user intent\cite{ref7}. Domain-specific knowledge bases hold substantial potential for knowledge management and reasoning\cite{ref8}.However, constructing such knowledge bases usually relies on web crawling, data generation, or existing industry databases\cite{ref2}, which can lead to issues with data accuracy and consistency.

Some recent studies explore methods for constructing domain-specific datasets. Sarasúa et al. discussed using large models to automatically generate datasets of any scale and domain\cite{ref9}, although this approach heavily relies on the prior knowledge embedded within the model. Addi et al. filtered higher-confidence data based on data attributes (such as the number of stars on open-source projects)\cite{ref10}, which limits applicability and may not accurately reflect true data quality. Apple’s Alex Fang and colleagues proposed using deep learning models, specifically DFNs, to filter image-text datasets to improve the quality of the data set\cite{ref11}. However, this approach may cause general embedding similarity methods to fail to capture semantic meaning as effectively as large models, and updates may lag.

\begin{figure}[!t]
\centering
\includegraphics[width=2.5in]{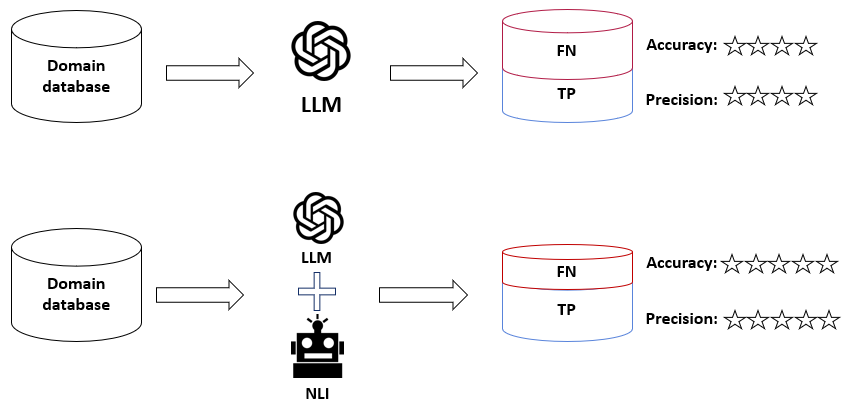}
\captionsetup{justification=raggedright, singlelinecheck=false}
\caption{Comparison of traditional filtering and the proposed self-nli-tdf framework. The robot icon represents pre-trained filtering models (e.g., large language models), and the assistant icon represents NLI (Natural Language Inference) models.
\newline
(a) Basic Filtering: This method uses pre-trained models to filter domain-specific data, relying on prior knowledge embedded in the model. However, it struggles with domain data that deviates from the training distribution, leading to higher false negatives (FN) and reduced filtering accuracy.
\newline
(b) self-nli-tdf Filtering: This iterative method leverages domain knowledge interrelations and combines reliable domain data with the model's internal knowledge for dual error detection. It effectively removes incorrect data by exploiting internal knowledge correlations.
\newline
Experimental Results: Compared to basic filtering, self-nli-tdf improves accuracy by reducing false negatives (red section in the figure)  Overall, the framework achieves better accuracy through dual detection.}
% \caption{Simulation results for the network.}
\label{fig_1}

\end{figure}

Correlations within domain-specific data are often overlooked in current filtering research. In response, we focused on the characteristics of domain datasets—specifically the interconnections between internal knowledge, and proposed the self-Natural Language Inference-trust data filter (self-nli-TDF) framework. This framework combines the generalization capabilities of large language models with NLI models specialized in NLI reasoning. Using internal prior knowledge of large language models, the framework performs reliability filtering for domain datasets.\cite{ref12} Additionally, it strengthens filtering performance by comparing trusted filtered knowledge with data awaiting filtering to deduce inferential relationships between them\cite{ref13}. The core of this framework lies in its iterative approach: initially filtering out data with relatively high reliability, then comparing trusted filtered knowledge with unfiltered data to deduce their inferential relationships, ultimately enhancing filtering accuracy and consistency.Figure.\ref{fig_1}

% 本图片展示了框架的改进点

Our contributions are as follows.

1. Introduction of the self-nli-tdf domain-specific data filtering framework: This paper proposes an iterative filtering approach by exploring the interrelationships within domain-specific knowledge bases, combined with high-confidence knowledge from the domain and the internal prior data of large models, creating a dual error detection concept. This approach effectively leverages internal knowledge associations within the domain and the model's prior information, resulting in a 1\%-5\% accuracy improvement over direct filtering in our dataset.

2.Mathematical Modeling of Knowledge Reliability: This paper introduces a mathematical model for quantifying the reliability of knowledge within knowledge bases, providing a more precise theoretical foundation for knowledge validation and data filtering, enhancing the scientific rigor of data filtering.

3.A Multi-Level Data Filtering Framework Integrating General and Domain-Specific Models: This study combines general large-scale language models with domain-specific models to construct an efficient data filtering framework. By merging the broad language understanding capabilities of general models with the specialized reasoning capabilities of domain models, a multi-level filtering structure is created. Additionally, a decision tree mechanism is introduced to balance conflicts and maintain consistency between model decisions, achieving more accurate data filtering results.

4.Open source code framework and domain-specific test datasets
As part of our contributions, we have open-sourced the \textbf{TrustDataFilter} framework and three domain-specific datasets. These resources provide researchers and practitioners with a robust toolkit for developing and evaluating data filtering approaches. The details are as follows:

\begin{itemize}
    \item \textbf{TrustDataFilter Framework:}  
    \url{https://github.com/littletigher/NLITrustDataFilter.git}

    \item \textbf{Radiation Domain Dataset:}  
    \url{https://huggingface.co/datasets/ZhJiHo/radiation_domain_dataset}  
    
    \textbf{Data Size:} 77.9k instances  
    
    \textbf{Last Updated:} November 15, 2024  

    \item \textbf{Science Domain Dataset:}  
    \url{https://huggingface.co/datasets/ZhJiHo/science_domain_dataset}  
    
    \textbf{Data Size:} 20.3k instances  
    
    \textbf{Last Updated:} November 15, 2024  

    \item \textbf{Biological Domain Dataset:}  
    \url{https://huggingface.co/datasets/ZhJiHo/biological_domain_dataset}  
    
    \textbf{Data Size:} 44.7k instances  
    
    \textbf{Last Updated:} November 15, 2024  
\end{itemize}

By offering the TrustDataFilter framework alongside these datasets, we aim to support the development and evaluation of innovative data filtering techniques across diverse fields, ensuring high-quality, domain-specific knowledge management and filtering solutions.

\section{Problem Formulation}

The study of human judgment in evaluating the correctness of domain-specific problems suggests that humans approach problem solving from multiple perspectives. First, they determine whether the description aligns with common sense \cite{ref20}, leveraging prior knowledge within their worldview to assess the issue. Second, they refer to relevant domain-specific materials to derive whether the problem can be inferred from known knowledge. Finally, they synthesize these two results, applying heuristic methods to arrive at the final judgment \cite{ref21}. This approach provides inspiration for constructing the Self-NLI framework to filter domain-specific data.

This perspective also highlights some limitations of existing domain data filtering frameworks:

\begin{enumerate}
    \item \textbf{Local Perspective:} Using large language models for data filtering is analogous to the first mode of reasoning, where prior knowledge within the worldview of the model is applied to assess the problem. On the other hand, methods based on deep learning or machine learning training resemble the second mode of reasoning, relying on past training data to draw conclusions. These approaches fail to integrate subjective reasoning with objective knowledge effectively.
    
    \item \textbf{Global Perspective Deficiency:} Current filtering frameworks often lack a mechanism to fully combine subjective reasoning and objective knowledge validation comprehensively. They fail to effectively integrate the prior knowledge of the model with factual knowledge within the domain. In practice, combining subjective judgment (i.e. common sense reasoning) with objective verification (i.e., logical inference based on external knowledge) can significantly enhance the accuracy and robustness of data filtering. However, existing frameworks fall short in this regard.
\end{enumerate}

To address these shortcomings, we propose the Self-NLI data filtering framework. This framework incorporates a Confidence Evaluation module, which utilizes the prior knowledge of large language models to determine whether the data to be filtered are trustworthy. It then applies a Contradiction Evaluation module, using an NLI model to assess whether data can be inferred from known trusted knowledge. Finally, a Decision Evaluation module is employed to make the final filtering judgement. The initial known trusted knowledge, along with the data deemed correct and trustworthy after filtering, is stored in a vector knowledge base as memory. As the filtering process progresses, this memory becomes increasingly enriched, thereby improving filtering performance within the given domain \cite{ref22}.
\subsection{Confidence Evaluation}

The Confidence Evaluation filter \( M(L, x, p) \) is a function based on an external large language model \( L \), where \( x \) represents the data to be evaluated, and \( p \) is a prompt containing filtering rules or output formats. The model \( L \), equipped with prior knowledge and reasoning capabilities, performs a reliability evaluation on the input data and outputs a reliability label \( Y \in \{0, 1\} \) along with a confidence score \( C \in [0, 1] \). Here, \( Y = 1 \) indicates that the data are reliable, while \( Y = 0 \) indicates that the data is unreliable. The confidence score \( C \) reflects the self-assessed certainty of the model about the outcome of the project. As discussed in recent studies\cite{ref25}, the probability of the model’s output label \( Y \) is a well-established approach to evaluating confidence, as it captures the uncertainty of the model in generating predictions.

For a given input \( x \) and prompt \( p \), the Confidence Evaluation filter outputs the reliability label \( Y \) and the confidence score \( C \) as follows:
\[
M(L, x, p) = (Y, C),
\]
where:
\begin{itemize}
    \item \( Y \in \{0, 1\} \): The reliability result determined by the large language model based on the input data \( x \) and the prompt \( p \).
    \item \( C \in [0, 1] \): The model's confidence score reflecting its certainty in the judgment result \( Y \).
\end{itemize}

\noindent Unlike traditional methods that maximize classification probabilities, both \( Y \) and \( C \) are obtained through the model's reasoning process, leveraging its prior knowledge and reasoning abilities. Specifically:
\begin{enumerate}
    \item When \( Y = 1 \), the confidence score \( C \) reflects the certainty of the model in the data being reliable.
    \item When \( Y = 0 \), the confidence score \( C \) reflects the certainty of the model in the data being unreliable.
\end{enumerate}

By allowing the large language model to perform the evaluation, the Confidence Evaluation filter integrates the model's prior knowledge and reasoning to determine the reliability of input data effectively.

\subsection{Contradiction Evaluation}

The contradiction evaluation \( N(I, x, x') \) is a function based on an external NLI (Natural Language Inference)\cite{ref23} model \( I \), where:
\begin{itemize}
    \item \( x \): The data to be evaluated.
    \item \( x' \): The most similar known trusted data retrieved from the vector database based on cosine similarity.
    \item \( I \): A pre-trained NLI model that determines the logical relationship between \( x \) and \( x' \).
\end{itemize}

Given the input \( x \) and \( x' \), the NLI model \( I \) outputs a relationship label \( Y \in \{-1, 0, 1\} \) and a confidence score \( C \in [0, 1] \). The definition of the label \( Y \) follows the work of Samuel R. Bowman et al. \cite{ref22}, where:
\begin{itemize}
    \item \( Y = 1 \): \textbf{Entailment}, which means \( x' \) logically implies \( x \).
    \item \( Y = 0 \): \textbf{Contradiction}, meaning \( x \) contradicts \( x' \).
    \item \( Y = -1 \): \textbf{Neutral}, meaning \( x \) and \( x' \) are unrelated.
\end{itemize}

The NLI model computes the conditional probability distribution for the three possible relationships:
\[
p_\theta(Y = 1 \mid x, x'), \quad p_\theta(Y = 0 \mid x, x'), \quad p_\theta(Y = -1 \mid x, x'),
\]
where \( p_\theta \) represents the conditional probability distribution parameterized by \( \theta \). The final relationship label \( y_2 \) is determined by selecting the label with the highest probability:
\[
Y = \arg\max_{k \in \{-1, 0, 1\}} p_\theta(Y = k \mid x, x').
\]

The confidence score \( C \) is defined as the maximum probability corresponding to the selected label \( Y \), expressed as:
\[
C = \max_{k \in \{-1, 0, 1\}} p_\theta(Y = k \mid x, x').
\]

Thus, the output of the contradiction filter \( N(I, x, x') \) is represented as:
\[
N(I, x, x') = (Y, C),
\]
where \( Y \) is the relationship label between \( x \) and \( x' \), and \( C \) is the confidence score reflecting the NLI model's certainty in the classification result.

\subsection{Decision Evaluation}

The Decision Evaluation model \( D(y_1, c_1, y_2, c_2) \) is based on a pre-trained decision tree\cite{ref24} \( T \), where \( y_1 \in \{0, 1\} \) represents the result from the Confidence Evaluation module, with \( c_1 \in [0, 1] \) being the associated confidence score. Similarly, \( y_2 \in \{-1, 0, 1\} \) represents the result from the Contradiction Evaluation module, with \( c_2 \in [0, 1\} \) being the corresponding confidence score. Here, \( y_2 = 1 \) indicates an entailment relationship, \( y_2 = 0 \) indicates a contradiction relationship, and \( y_2 = -1 \) indicates a neutral relationship.

Given the input feature vector \(\mathbf{x} = (y_1, c_1, y_2, c_2)\), the pre-trained decision tree model \( T \) in the Decision Evaluation module performs a comprehensive judgment and outputs the final result \( z \in \{0, 1\} \). Specifically, \( z = 1 \) indicates that the knowledge is deemed reliable, while \( z = 0 \) indicates that the knowledge is unreliable. The decision tree automatically learns the mapping relationship between the input features and the final decision through pre-training.

The final output of the Decision Evaluation model is expressed as:
\[
z = D(y_1, c_1, y_2, c_2) = T(\mathbf{x}),
\]
where \( \mathbf{x} = (y_1, c_1, y_2, c_2) \) is the input feature vector, \( T \) is the pre-trained decision tree, and \( z \) is the final decision result indicating the reliability of the knowledge.

Conceptually, Self-NLI, as a domain-specific data filtering framework based on large language models and natural language inference models, offers several advantages:

\textbf{Generality}: Self-NLI combines the prior knowledge of large language models with the logical reasoning of natural language inference, enabling flexible application to various domain-specific data filtering tasks.

\textbf{Modularity}: The framework's Confidence Evaluation, Contradiction Evaluation, and Decision Evaluation modules can be independently replaced and optimized, providing a high degree of modularity.

\textbf{Convenience}: The framework does not rely on extensive model training; it only requires pre-trained large language models and natural language inference models to perform data filtering tasks effectively.

The next section will provide a detailed introduction to the architecture and processing flow of the Self-NLI framework.

\section{Methodology}
In this chapter, (1) we will introduce the static architecture of the self-NLI-DataFilter, outlining its fundamental components and their relationships.(2) we will describe the dynamic process, illustrating how the system operates to filter and process data step by step.
(3) we will elaborate on the key design aspects of the workflow, breaking down the process into four critical steps:

Trusted Knowledge Matching

Contradiction Evaluation

Confidence Evaluation

Decision Tree Evaluation
\subsection{overall architecture}
\begin{figure}[!t]
\centering
\includegraphics[width=0.5\textwidth]{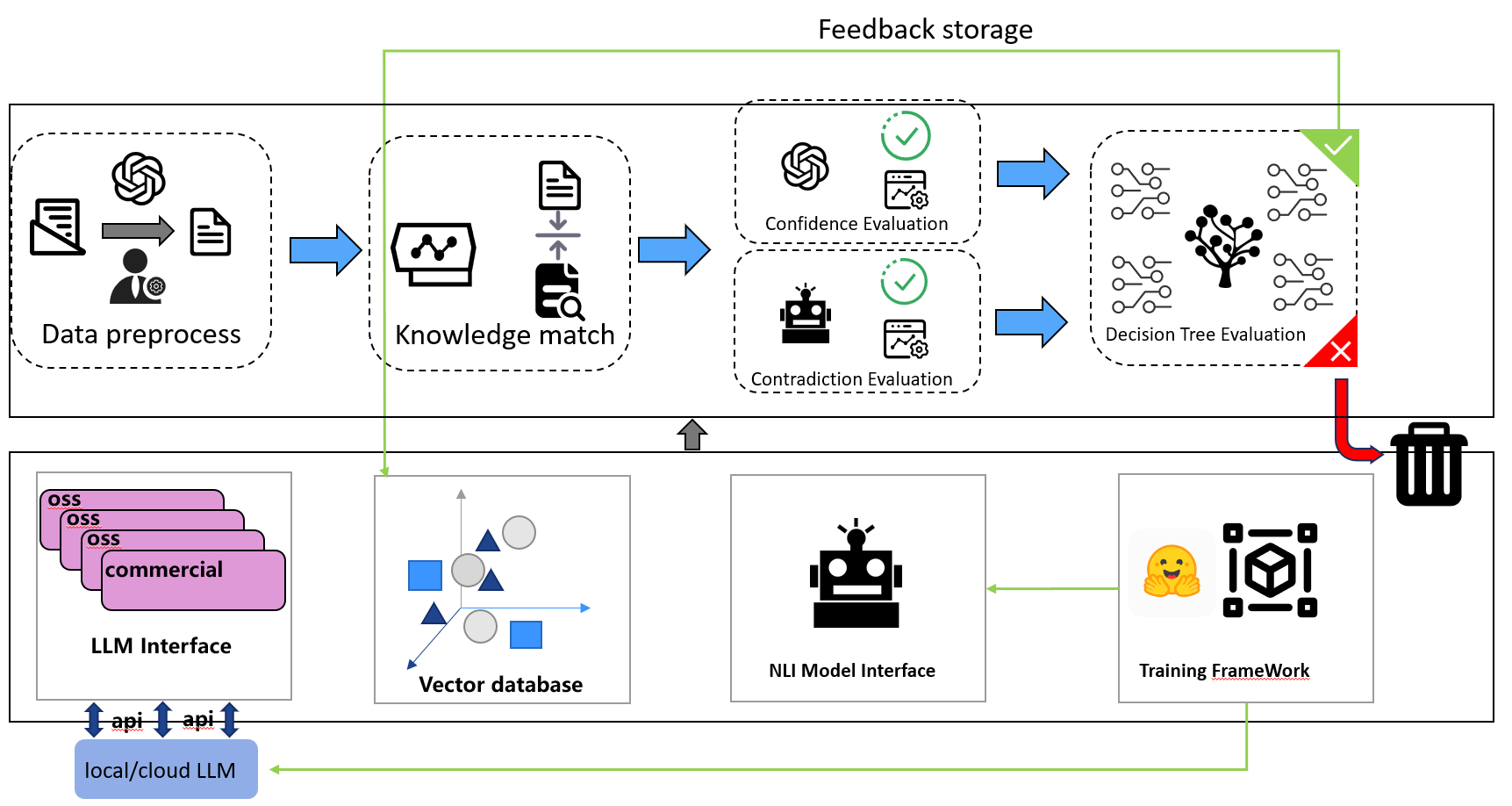}
\captionsetup{justification=raggedright, singlelinecheck=false}
\caption{self-nli-datafilter architecture}
% \caption{Simulation results for the network.}
\label{fig_2}

\end{figure}

Our complete system architecture is shown in Figure \ref{fig_2}. It is structured into two main layers: the business layer and the foundational layer, working together to achieve efficient and accurate data filtering.

\textbf{Business Layer:}  
The top layer represents the data filtering process, which encompasses steps from preprocessing to final storage. Initially, 5\% of the data is selected for annotation to train both the NLI model and the decision tree. Inspired by Van Engelen and Hoos \cite{ref26}, this approach demonstrates that utilizing a small annotated subset can effectively enhance model performance, a common practice in data processing \cite{ref17}. The remaining data are passed to the Knowledge Match module, which performs vector similarity matching between the data and the trusted knowledge base. The matched items are then processed by the Confidence and Contradiction Evaluation module, consisting of two steps: confidence evaluation and contradiction evaluation. Confidence evaluation assesses the reliability of the data, while contradiction evaluation determines whether the data conflict with the trusted knowledge in the knowledge base. The results of both evaluations are passed to the Decision Tree Evaluation module, which performs a comprehensive evaluation to determine the trustworthiness of the data. Trusted data is stored in the knowledge base, while untrustworthy data is discarded, ensuring the accuracy of the filtering process.

\textbf{Foundational Layer:}  
The bottom layer provides the technical backbone for the system, comprising the training framework, vector database, and interfaces for large language models (LLM) and NLI models.The training framework leverages Hugging Face for pre-training and fine-tuning, ensuring the robustness and accuracy of the NLI model and decision tree. The LLM and NLI interfaces are implemented using LangChain, enabling seamless integration for tasks like semantic analysis and contradiction evaluation. The vector database, powered by Milvus, efficiently stores and manages vectorized knowledge data, supporting effective knowledge matching and filtering operations. Together, these components form the foundation that supports the business layer, ensuring a reliable and efficient workflow for data filtering.

\subsection{overall process}
\begin{figure}[!t]
\centering
\includegraphics[width=0.5\textwidth]{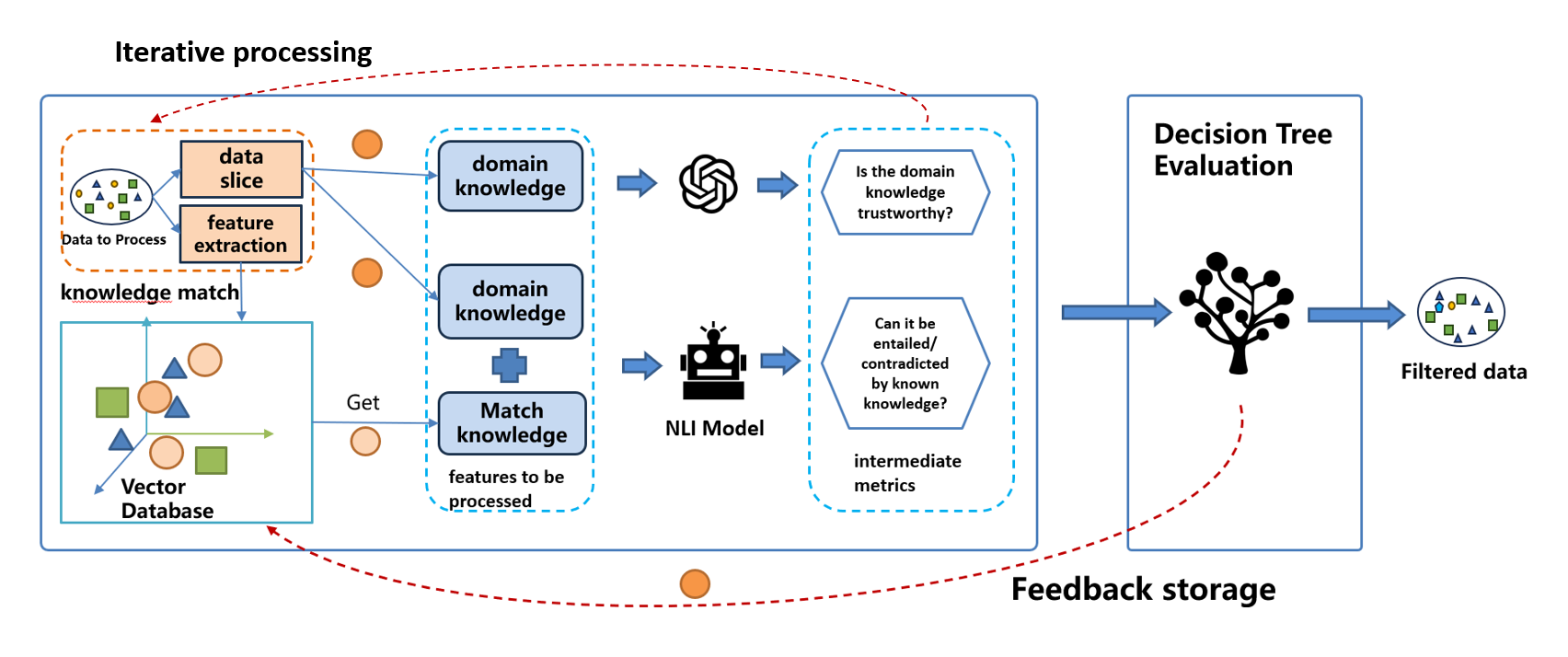}
\captionsetup{justification=raggedright, singlelinecheck=false}
\caption{self-nli-datafilter architecture}
% \caption{Simulation results for the network.}
\label{fig_3}
\end{figure}

Our frame process is shown in Figure \ref{fig_3}. It consists of a systematic sequence of stages designed to ensure accurate and efficient knowledge filtering.

\textbf{Overall Process:}
The process begins with the Data to Process stage, where each knowledge item undergoes initial filtering. Subsequently, the knowledge items are passed to the Trusted Knowledge Matching stage, where they are matched with trusted items in the vector knowledge base. If a match is successful, the knowledge item proceeds for further evaluation. If the match fails, the item is temporarily stored in the Process in Next Iteration stage for re-evaluation in subsequent iterations.

\textbf{Detailed Steps:}
For items with a successful match, a confidence evaluation and contradiction evaluation are conducted. These evaluations utilize a large language model and an NLI model to assign confidence and contradiction flags along with their corresponding scores. The results of these evaluations are then integrated and passed to the Decision Tree Evaluation stage, where the final outcome is determined. Knowledge items that pass the evaluation are marked as qualified and stored in the knowledge base. Items that do not meet the criteria are marked as not qualified and sent to the discard stage. The process continues until all knowledge items are processed, concluding at the "End" stage.

\textbf{Iterative Improvements:}
Our framework draws inspiration from certain aspects of the retrieval-augmented generation (RAG) framework, particularly the concept of matching each piece of knowledge with the most relevant reference entry from the vector database \cite{ref18}. However, unlike RAG, which primarily uses matching for prompt construction, our framework focuses on assisting knowledge filtering through natural language inference reasoning. As highlighted by Barnett et al. \cite{ref27}, vector matching accuracy is a significant challenge in RAG systems. Semantic misalignment between queries and indexed vectors can lead to retrieval failures or low-quality results, thereby affecting overall system performance.

To address these challenges, we propose an iterative approach to handle unmatched data. Unmatched items are deferred to subsequent iterations for re-evaluation. This iterative process incrementally improves the quality of the vector database by integrating validated knowledge, which in turn enhances matching performance over time. In contrast, standard RAG frameworks typically address unmatched references by either generating a default response or providing a placeholder such as "I am unable to process this query" \cite{ref19}.

\subsection{Trusted Knowledge Matching}
The process of aligning data with a trusted knowledge base is fundamental to effective contradiction evaluation. This process is organized into three critical components: knowledge base construction, vector matching techniques, and dynamic knowledge base expansion.

\textbf{Knowledge Base Construction:}
To begin, we construct a domain-specific knowledge base containing validated and reliable information. For domains with a sufficient amount of trustworthy data, this data is directly vectorized and stored in the database. In specialized domains with limited data availability, 5\% of the data is manually marked as trustworthy and incorporated into the knowledge base during filtering. This ensures that even in data-scarce domains, the knowledge base provides a reliable foundation for subsequent filtering and evaluation.

\textbf{Vector Matching Techniques:}
Our vector matching is based on the DPR method \cite{ref28}, which employs a Dual-Encoder Architecture to independently encode queries and documents into dense embeddings for fast and precise semantic matching. The sentence-transformers/all-MiniLM-L6-v2 model is utilized for vectorization, while the IVF\_FLAT indexing technique is used to enhance retrieval efficiency. Cosine similarity is applied to calculate vector similarity, focusing on semantic alignment between items. A predefined similarity threshold, typically between 0.85 and 0.90, ensures that only highly relevant data items are matched. Items falling below this threshold are marked as "unmatched" and temporarily stored for reassessment in subsequent filtering cycles. As related studies highlight \cite{ref29}, cosine similarity does not strictly measure semantic distance, so the threshold is heuristically assigned based on empirical observations.

\textbf{Dynamic Knowledge Base Expansion:}
This iterative approach enables continuous refinement of the knowledge base by integrating validated data from each filtering cycle. Unmatched data items are reevaluated in subsequent iterations, allowing for gradual accumulation of trustworthy information. Newly validated data is vectorized and added to the knowledge base, improving its semantic coverage and matching performance over time. Unlike static approaches, this strategy actively adapts the knowledge base to changing domain requirements and enhances its robustness and scalability. The iterative process ensures that the filtering mechanism evolves with each cycle, leveraging newly added reliable knowledge to improve subsequent filtering precision.

By integrating rigorous knowledge base construction, advanced vector matching techniques, and an iterative expansion process, this framework maintains the integrity of the knowledge base. It prevents irrelevant or insufficiently similar data from being misclassified as trustworthy, enabling more accurate contradiction detection in the evaluation phase. This adaptive strategy provides a robust foundation for a trusted knowledge filtering pipeline, ensuring it meets both current and future requirements.

\subsection{Contradiction Evaluation}
Contradiction evaluation is a key process that ensures the consistency and reliability of the knowledge base. This process is structured into three main aspects: its mechanism, outputs, and role in the broader data filtering framework.

\textbf{Mechanism of Contradiction Evaluation:}
Contradiction evaluation assesses the MNLI (Multi-Genre Natural Language Inference) \cite{ref30} relationship between the data to be filtered and the existing trusted knowledge using a pre-trained model. After similarity matching, the system forwards the data to be filtered (referred to as knowledge) and the matched knowledge (matchKnowledge) to the contradiction evaluation module. This module uses a RoBERTa-MNLI model fine-tuned in the MNLI domain to infer the relationship between the two:

If the inference result indicates that the trusted knowledge entails the data to be filtered, the information in the data is considered relatively credible.
If the result shows a contradiction, the knowledge contained in the data is likely erroneous.
When the result is neutral, it suggests the knowledge in the data is new or unrelated, making it infeasible to assess credibility based solely on contradiction evaluation.
Outputs of Contradiction Evaluation:
The system produces two final outputs: contradict\_flag, which indicates the inference result, and score\_of\_contradict, which provides the confidence score. These outputs enable the system to identify data that conflicts with trusted knowledge while retaining novel information that lacks inherent contradictions. This ensures that only consistent data is added to the knowledge base, maintaining its integrity while allowing for dynamic expansion.

\textbf{Role in Data Filtering Framework:}
In the broader context of data filtering, contradiction evaluation acts as a crucial gatekeeping mechanism. By identifying and excluding contradictory data, it safeguards the consistency and accuracy of the knowledge base. This process also enables the incremental inclusion of new, potentially valuable information, which may be further evaluated and validated. By systematically incorporating contradiction assessment into the filtering architecture, the framework enhances the reliability of data inclusion, ensuring high-quality knowledge management and enabling intelligent, data-driven decision support.

Through contradiction evaluation, the system effectively maintains the consistency of the knowledge base by filtering out contradictory data while retaining novel, non-conflicting information. This filtering mechanism ensures that the knowledge base remains accurate and adaptable, providing a solid foundation for efficient knowledge management and dependable decision-making.

\subsection{Confidence Evaluation}
Confidence evaluation plays a critical role in determining the reliability of data during filtering, ensuring that only trustworthy information is integrated into the knowledge base. This process is organized into three key aspects: its mechanism, implementation, and role within the data filtering framework.

\textbf{Mechanism of Confidence Evaluation:}
Confidence evaluation assesses the credibility of data by generating two key outputs for each data item: the confidence flag  and the confidence score. The confidence flag is a binary indicator, where 0 signifies that the data is generally not trustworthy, and 1 signifies that the data is generally trustworthy. The confidence score quantifies the certainty of this judgment, with higher scores indicating a more reliable assessment. Together, these outputs provide a critical foundation for filtering decisions.

\textbf{Implementation of Confidence Evaluation:}
To perform confidence evaluation, the framework uses a large language model as the core component, leveraging a prompt-based framework to guide the model in generating confidence indicators. This step primarily relies on the extensive prior knowledge encoded in the pre-trained model to assess data reliability. Studies such as On Reliability and Robustness of Current Estimators \cite{ref31} have demonstrated the ability of large language models to effectively evaluate the factual reliability of their outputs. Additionally, research on MONITOR \cite{ref32} provides systematic approaches to assess the consistency of LLMs across various contexts, further supporting their application in confidence-based tasks. These findings underline the robustness of using pre-trained language models for confidence evaluation, enabling accurate and nuanced data filtering.

\textbf{Role in the Data Filtering Framework:}
Confidence evaluation measures the alignment of the data with known, validated knowledge or patterns, providing a quantitative basis for trustworthiness. Data with high confidence scores is deemed reliable and included in the knowledge base, while data with low scores is flagged for further analysis or excluded. By integrating confidence evaluation, the filtering framework gains an additional layer of quality control, ensuring that only high-confidence information is retained. This mechanism minimizes the risk of incorporating inaccurate or misleading data, significantly boosting the reliability and robustness of the system.

\subsection{Decision Tree Evaluation}
Decision Tree Evaluation serves as the final judgment step in the data filtering process, integrating the outcomes of Confidence Evaluation and Contradiction Evaluation to determine the overall trustworthiness of data. This process can be divided into three main aspects: its mechanism, integration of evaluation standards, and interpretability.

\textbf{Mechanism of Decision Tree Evaluation:}
Decision Tree Evaluation relies on the results of prior evaluations, including confidence\_flag, score\_of\_confidence, contradict\_flag, and score\_of\_contradict. These outputs, represented as integers (int) or floating-point numbers (float), are well suited for processing with a decision tree model. Decision trees are highly effective at handling numerical and categorical features, constructing hierarchical decision structures based on splitting rules such as information gain or Gini impurity. Studies have demonstrated that decision trees perform well on datasets with numerical features while maintaining high interpretability \cite{ref33}. The decision tree processes these inputs to output a final decision flag:
If the flag is 1, the data is deemed trustworthy and incorporated into the knowledge base.
If the flag is 0, the data is considered untrustworthy and discarded.
Integration of Evaluation Standards:
The decision tree seamlessly combines the outputs of Confidence Evaluation and Contradiction Evaluation to enable a comprehensive assessment. By incorporating confidence\_flag, score\_of\_confidence, contradict\_flag, and score\_of\_contradict as input variables, the decision tree flexibly performs hierarchical branching based on various conditions and weightings. This integration allows the model to balance the contributions of each evaluation: in some cases, relying more heavily on confidence scores, while in others prioritizing contradiction levels. For instance, when a data item has a high score\_of\_confidence but also a high score\_of\_contradict, the decision tree can recognize these conflicting signals and autonomously select the most appropriate path to output the final flag value. This multi-dimensional judgment mechanism minimizes the risk of misjudging data credibility based on any single standard, ensuring accurate data filtering and supporting knowledge base quality management and expansion.

\textbf{Interpretability of Decision Tree Evaluation:}
A key strength of the decision tree is its interpretability. During data filtering, the decision path generated by the model reveals the reasoning behind each step, rendering the judgment process transparent. By examining the branching structure, users can clearly understand how various judgment conditions influence the final decision. This transparency facilitates analysis and validation, allowing users to identify potential issues or optimization opportunities within the filtering process more effectively. The decision tree’s logical structure thus provides a clear and interpretable standard for determining data reliability, enhancing the precision and reliability of the overall data filtering process.

\textbf{Conclusion:}
Through its hierarchical structure and flexibility, Decision Tree Evaluation integrates multiple evaluation conditions to provide a robust and accurate judgment mechanism. It ensures that only trustworthy data is included in the knowledge base while maintaining a transparent decision-making process. This enhances the quality and reliability of data filtering, providing critical support for knowledge base management and dynamic expansion.

\section{DataSet Construct}
To demonstrate the effectiveness of self-NLI-DataFilter, our experiments require domain-specific datasets for filtering. However, not all domain datasets are publicly available. In this study, to validate the algorithm's performance, we constructed three experimental domain datasets using three different methods. Each record in the constructed domain datasets consists of a knowledge statement and a flag. The knowledge represents a statement of a knowledge point within the domain, while the flag serves as an indicator, where 0 denotes incorrect knowledge and 1 denotes correct knowledge.The three datasets we constructed are as follows, and their composition is introduced below:

Biological Dataset:
This dataset consists of biological knowledge generated by a large language model under prompt engineering.We adopted the self-instruct\cite{ref34} approach for data generation. The data includes both correct and incorrect knowledge points based on the model's prior knowledge.

Science Dataset:
This dataset was constructed by extracting knowledge point descriptions from the scientific knowledge base "allenai/ai2\_arc" Datasets\cite{ref35} at Hugging Face using a large language model. The ai2\_arc dataset comprises 7,787 elementary-level multiple-choice science questions. The large model was utilized to extract correct and incorrect knowledge descriptions to form this dataset.

Radiation Dataset:  
This dataset contains data related to nuclear radiation protection. We collected content from websites, including CBRNResponder, the U.S. Environmental Protection Agency (EPA), FEMA, and the IAEA (International Atomic Energy Agency), using web crawlers. Additionally, we included academic content from Google Scholar and ScienceDirect.Based on the capabilities of large language models in information extraction \cite{ref36}, we used a large language model to organize and refine the domain-specific knowledge points from the collected data. The collected data was segmented and processed to extract knowledge points, forming the dataset.

\subsection{Biological Dataset}
The Biological dataset was constructed to provide a comprehensive and diverse foundation for domain-specific data filtering experiments. This process can be divided into three key aspects: dataset construction, data generation techniques, and the dataset's advantages.

\begin{figure}[!t]
\centering
\includegraphics[width=2.5in]{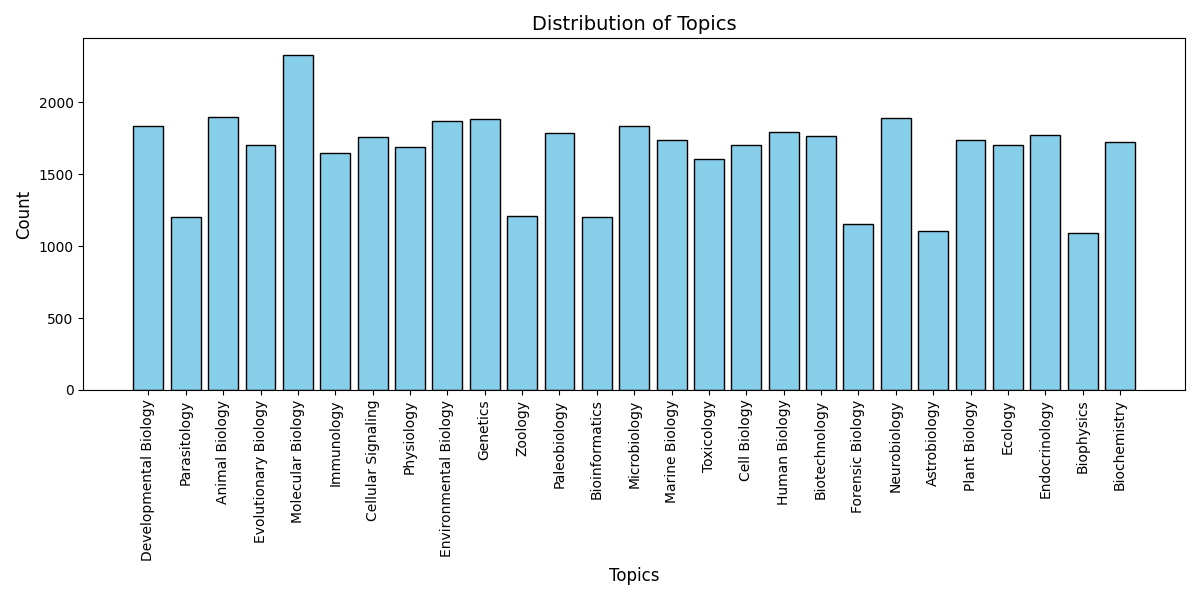}
\captionsetup{justification=raggedright, singlelinecheck=false}
\caption{Distribution of Topics}
% \caption{Simulation results for the network.}
\label{fig_4}
\end{figure}

\textbf{Dataset Construction:}
The dataset was created using GPT-3.5 based on its prior knowledge. We implemented and modified the self-instruction framework to enable the large language model to generate both correct and incorrect biological knowledge points based on given topics and subtopics. Through a heuristic approach, we pre-constructed 27 topics (Figure \ref{fig_4}), each containing several subtopics, resulting in a total of over 200 knowledge generation themes.

\textbf{Data Generation Techniques:}
During the data generation process, we iteratively traversed all combinations of topics and subtopics and employed prompt engineering techniques. Carefully designed prompts guided the model to generate biological knowledge statements that included both accurate and inaccurate knowledge points. This iterative process ensured both diversity and richness in the generated data. In total, 44.7k knowledge points were generated. The dataset has been open-sourced on Hugging Face under the repository ZhJiHo/biological\_domain\_dataset.

\textbf{Advantages of the Dataset:}
This data generation method provides the Biological dataset with several notable strengths. First, the pre-designed 27 topics and over 200 subtopics ensure broad and diverse content, covering themes ranging from macro-level ecology to molecular biology. This breadth provides a reliable foundation for comprehensive filtering experiments within the biological domain. Second, the generation of incorrect knowledge points involved an intentional distortion of correct knowledge points. By instructing the model to first produce accurate knowledge and then deliberately creating incorrect versions, the dataset ensures that the incorrect knowledge points are sufficiently deceptive. This makes the dataset particularly valuable for testing the reliability and robustness of filtering models under challenging conditions.

\subsection{Science Dataset}
The Science dataset was constructed to provide a structured and high-quality corpus for domain-specific experiments in the scientific field. This process can be divided into three key aspects: dataset construction, methodology, and dataset significance.

\textbf{Dataset Construction:}
The Science dataset was created by extracting knowledge point descriptions from the scientific knowledge base ``allenai/ai2\_arc'' Datasets at Hugging Face. This source contains multiple elementary-level multiple-choice science question datasets. For each question, we combined the question with the correct option to form a correct answer and used a large language model to extract accurate knowledge points from the answer. Similarly, we combined the question with incorrect options to create incorrect answers, tasking the large language model with generating inaccurate knowledge points based on these incorrect answers. Through this process, we transformed the original multiple-choice questions into a domain-specific dataset with pre-labeled correct and incorrect descriptions. In total, 20.3k knowledge points were generated, which we have open-sourced on Hugging Face under the repository ZhJiHo/science\_domain\_dataset.

\textbf{Methodology:}
This method effectively leverages validated knowledge bases and transforms their high-quality information into structured scientific domain data. By systematically combining questions and options, the dataset includes both correct and incorrect knowledge points, ensuring diversity while maintaining semantic and logical clarity. The large language model plays a key role in extracting precise knowledge from correct answers and generating inaccurate knowledge points from incorrect answers, adding authenticity and complexity to the dataset.

\textbf{Dataset Significance:}
The resulting dataset provides a rich and challenging corpus for domain-specific experiments. It supports comprehensive evaluations of knowledge filtering frameworks and related methodologies by offering pre-labeled data with high semantic clarity and logical consistency. The diversity and complexity of the dataset make it particularly valuable for testing filtering models under realistic and demanding conditions, enhancing its utility for advancing domain-specific knowledge management.

\subsection{Radiation Dataset}
The Radiation domain dataset was constructed to provide a challenging and authentic resource for evaluating domain-specific knowledge filtering frameworks. This process can be divided into three main aspects: data collection, data processing, and the dataset's purpose and significance.

\textbf{Data Collection:}
The dataset was primarily collected by our researchers through web scraping. Initially, radiation-related data was crawled from authoritative official websites, including CBRNResponder, the U.S. Environmental Protection Agency (EPA), FEMA, and the International Atomic Energy Agency (IAEA). Due to the limited volume of data from these sources, additional radiation-related content was scraped from academic papers available on Google Scholar and ScienceDirect. Combining data from both official sources and academic literature resulted in the construction of an initial web-scraped domain dataset. Since the original data came from authoritative sources, we assumed the extracted knowledge to be accurate. Furthermore, the reliance on web scraping provides the dataset with a high degree of authenticity and real-world relevance.

\textbf{Data Processing:}
In the data processing phase, the scraped text was filtered and segmented into smaller text blocks. A large language model was then employed to extract structured knowledge points from these text slices, transforming raw text into usable knowledge entries. To enrich the dataset, we generated inaccurate knowledge points based on the correct knowledge points, creating a dataset with both accurate and inaccurate entries. Through this process, we constructed approximately 77.9k radiation domain knowledge points, which we have open-sourced on Hugging Face under the repository ZhJiHo/radiation\_domain\_dataset.

\textbf{Purpose and Significance:}
The Radiation domain was deliberately chosen as a relatively niche area. Subsequent experiments revealed a noticeable decline in model accuracy within this domain, highlighting the model's limited prior knowledge of radiation-related topics. The primary goal of constructing this dataset is to evaluate the extent to which the self-NLI-DataFilter framework can improve model performance when prior domain knowledge is insufficient. By introducing this specialized dataset, we created a challenging test environment to comprehensively assess the framework's ability to filter and evaluate domain-specific knowledge. This approach not only measures the robustness and adaptability of the framework but also evaluates its potential to enhance performance when supplemented with external domain knowledge. Additionally, it lays a solid foundation for developing more efficient domain-specific knowledge management systems.

\section{Experiment}
{\bf{Experiment Setup:}}We prepared three domain-specific datasets: biological, science, and radiation, along with three large language models: RoBERTa, Qwen2, and GPT-3.5. Each dataset was divided into 5\% for training, 5\% for validation, and 90\% for testing. The experiments utilized three distinct filtering methods:

\begin{itemize}
    \item \textbf{Basic Filtering}: This method directly uses the large language models to perform filtering without any additional framework or enhancements.
    \item \textbf{Self-NLI Filtering}: This method leverages the framework proposed in this paper, integrating self-NLI (Natural Language Inference) processes for filtering.
    \item \textbf{Fake Filtering}: This method applies an ablation experiment, where the trusted knowledge matching in the proposed Self-NLI framework is replaced with random irrelevant knowledge matching.
\end{itemize}

Each model applied these filtering approaches to the 90\% test portion of each dataset, and the results were systematically evaluated and compared.

{\bf{Datasets:}} We constructed datasets from three distinct domains: biological, science, and radiation, to serve as benchmarks for evaluating domain-specific data filtering tasks. The biological dataset was generated through self-reasoning by large language models, ensuring both domain relevance and diversity. The science dataset was derived from publicly available open-source datasets, representing widely accepted and standardized knowledge in the scientific domain. Finally, the radiation dataset was collected using web scraping techniques, capturing domain-specific data from real-world scenarios.

{\bf{Model.}} 
We utilized three public models: RoBERTa (Liu, Y \cite{ref4}), Qwen2-7B (Yang, An, et al \cite{ref5}), and GPT-3.5 API. Since the RoBERTa base model cannot directly perform data filtering, we fine-tuned it using 5\% of the dataset, as mentioned earlier, as the training set and another 5\% as the validation set. The other large language models, Qwen2-7B and GPT-3.5, were used in their original, untrained state.

\subsection{Experimental Procedure}
We conducted experiments to evaluate dataset filtering performance using three domain-specific datasets—biological, science, and radiation—and three large language models: RoBERTa, Qwen2, and GPT-3.5. Each dataset was divided into 5\% for training, 5\% for validation, and 90\% for testing.And We test Three filtering methods.The experimental process consisted of the following steps:

\textbf{Step 1 Training the RoBERTa Model:}
Since the RoBERTa model cannot directly execute filtering tasks, it was first trained on the original datasets. The training datasets included \textit{biological-train}, \textit{science-train}, and \textit{radiation-train}. Correspondingly, the validation datasets used were \textit{biological-valid}, \textit{science-valid}, and \textit{radiation-valid}. The model’s performance on the validation datasets was evaluated after training to ensure its readiness for subsequent tasks.

\textbf{Step 2 Training the Decision Tree in the Self-NLI Framework:}
The Self-NLI framework incorporates a pre-trained decision tree model, which was trained on the same training datasets: \textit{biological-train}, \textit{science-train}, and \textit{radiation-train}. Similarly, the validation datasets used were \textit{biological-valid}, \textit{science-valid}, and \textit{radiation-valid}. The validation results were analyzed to assess the performance and accuracy of the decision tree.

\textbf{Step 3 Basic Filtering:}
In the basic filtering method, we referred to the study by Si et al.\cite{ref37} on large language models' ability to verify truthfulness. Using prompt-based techniques, we directly instructed the three large language models (RoBERTa, Qwen2, and GPT-3.5) to determine whether the data was trustworthy, and the results were recorded for further analysis and comparison.

\textbf{Step 4 Initializing the Self-NLI Framework:}
To initialize the Self-NLI framework, correct knowledge extracted from the training datasets (\textit{biological-train}, \textit{science-train}, and \textit{radiation-train}) was pre-stored in a vector database. This vectorized knowledge base served as the foundation for the framework, enabling efficient knowledge retrieval during filtering.

\textbf{Step 5 Self-NLI Filtering:} 
Using the Self-NLI filtering method, the same three models (RoBERTa, Qwen2, and GPT-3.5) were employed to filter the test datasets: \textit{biological-test}, \textit{science-test}, and \textit{radiation-test}. The filtering process iterated five times, and the results of each iteration were carefully recorded.

\textbf{Step 6 Fake Filtering:} 
For the ablation experiment, the Self-NLI framework was modified by replacing the trusted knowledge matching mechanism with random irrelevant knowledge matching.
Our fake-related data is randomly selected from the Facebook/ASSET dataset\cite{ref38}, which contains 4.5k simplified statements. These statements are concise, unrelated to our three domains, and suitable as a control group for the ablation experiments.This fake filtering method was applied to the same test datasets (\textit{biological-test}, \textit{science-test}, and \textit{radiation-test}) using RoBERTa, Qwen2, and GPT-3.5. The results of this step were analyzed and compared with those from basic filtering and Self-NLI filtering to evaluate the importance of accurate knowledge matching.

After completing the experiments and calculating the confusion matrix for each result, we used accuracy, precision, recall, and F1 score as evaluation metrics. Accuracy reflects the overall correctness of the model's predictions and is suitable for assessing global performance. Precision focuses on the accuracy of positive predictions, making it ideal for scenarios with strict requirements on false positives. Recall evaluates the coverage of positive samples, which is critical for tasks that prioritize minimizing omissions. The F1 score balances precision and recall, providing a more comprehensive reference.

\subsection{Experimental Result}

\begin{table*}[!t]
\caption{Experimental Results\label{tab:results}}
\centering
\resizebox{0.8\textwidth}{!}{
\begin{tabular}{lllcccc}
\toprule
\textbf{Dataset}    & \textbf{Model}  & \textbf{Methodology} & \textbf{Accuracy} & \textbf{Precision} & \textbf{Recall} & \textbf{F1 Score} \\ 
\midrule
\multirow{7}{*}{Biological} 
                    & \multirow{2}{*}{Qwen2}    & Base                 & 0.8863            & 0.8383             & 0.9914          & 0.9084            \\ 
                    &                          & Fake                 & 0.8937            & 0.8500             & 0.9873          & 0.9135            \\ 
                    &                          & Self-NLI             & 0.9030            & 0.8673             & 0.9797          & 0.9201            \\ 
\cmidrule{2-7}
                    & \multirow{2}{*}{GPT}     & Base                 & 0.8640            & 0.8120             & 0.9900          & 0.8920            \\ 
                    &                          & Fake                 & 0.8323            & 0.8774             & 0.8250          & 0.8504            \\ 
                    &                          & Self-NLI             & 0.8847            & 0.8713             & 0.9358          & 0.9024            \\ 
\cmidrule{2-7}
                    & \multirow{2}{*}{RoBERTa} & Base                 & 0.8799            & 0.8865             & 0.9045          & 0.8954            \\ 
                    &                          & Fake                 & 0.8525            & 0.8643             & 0.8786          & 0.8714            \\ 
                    &                          & Self-NLI             & 0.8831            & 0.9085             & 0.8833          & 0.8957            \\ 
\midrule
\multirow{7}{*}{Science} 
                    & \multirow{2}{*}{Qwen2}    & Base                 & 0.8953            & 0.9431             & 0.9296          & 0.9363            \\ 
                    &                          & Fake                 & 0.8900            & 0.9172             & 0.9532          & 0.9348            \\ 
                    &                          & Self-NLI             & 0.8986            & 0.9241             & 0.9561          & 0.9398            \\ 
\cmidrule{2-7}
                    & \multirow{2}{*}{GPT}     & Base                 & 0.9315            & 0.9393             & 0.9806          & 0.9595            \\ 
                    &                          & Fake                 & 0.9301            & 0.9439             & 0.9734          & 0.9584            \\ 
                    &                          & Self-NLI             & 0.9353            & 0.9481             & 0.9753          & 0.9615            \\ 
\cmidrule{2-7}
                    & \multirow{2}{*}{RoBERTa} & Base                 & 0.9287            & 0.9207             & 1.0000          & 0.9587            \\ 
                    &                          & Fake                 & 0.8917            & 0.9174             & 0.9552          & 0.9359            \\ 
                    &                          & Self-NLI             & 0.9653            & 0.9741             & 0.9843          & 0.9792            \\ 
\midrule
\multirow{7}{*}{Radiation} 
                    & \multirow{2}{*}{Qwen2}    & Base                 & 0.8376            & 0.9390             & 0.8685          & 0.9023            \\ 
                    &                          & Fake                 & 0.8833            & 0.9139             & 0.9560          & 0.9345            \\ 
                    &                          & Self-NLI             & 0.8885            & 0.9346             & 0.9375          & 0.9360            \\ 
\cmidrule{2-7}
                    & \multirow{2}{*}{GPT}     & Base                 & 0.8500            & 0.9400             & 0.8841          & 0.9112            \\ 
                    &                          & Fake                 & 0.8610            & 0.8746             & 0.9810          & 0.9247            \\ 
                    &                          & Self-NLI             & 0.9092            & 0.9348             & 0.9628          & 0.9486            \\ 
\cmidrule{2-7}
                    & \multirow{2}{*}{RoBERTa} & Base                 & 0.8780            & 0.8775             & 0.9999          & 0.9347            \\ 
                    &                          & Fake                 & 0.8588            & 0.8776             & 0.9735          & 0.9231            \\ 
                    &                          & Self-NLI             & 0.9482            & 0.9638             & 0.9774          & 0.9706            \\ 
\bottomrule
\end{tabular}
}
\end{table*}

Table \ref{tab:results}  presents a comparison of our method with existing baseline models across datasets in different domains, including biology, science, and radiation. On average, our method achieved a (3\%) improvement in accuracy, (2.6\%) in precision, (0.5\%) in recall, and (1.7\%) in F1 score, demonstrating the effectiveness of SelfNLI in enhancing the data filtering capability of models.

We find that the performance gains varied across datasets and models. For example:The highest improvement was observed with the Radiation-RoBERTa model\cite{ref4}, achieving a (7.02\%) increase in accuracy.The lowest improvement was with Biological-RoBERTa, which showed only a (0.32\%) increase in accuracy.

The following sections provide a detailed analysis of the observed improvements, focusing on both dataset-specific and model-specific performance patterns.

\textbf{Biological Dataset:}  
Although the overall performance improvement was relatively lower, the method demonstrated a notable increase in precision (2.12\%), making it well-suited for scenarios that require high accuracy and low false positives.

\textbf{Science Dataset:}  
SelfNLI exhibited the highest average accuracy improvement (3.90\%) among all datasets, indicating its superior capability in addressing the complex data filtering tasks required in scientific domains.

\textbf{Radiation Dataset:}  
The most significant improvement was observed in F1 score (2.51\%), highlighting the reliability of SelfNLI in processing diverse domain knowledge.

\textbf{RoBERTa:}  
Achieved the highest precision improvement (5.39\%), but with a slight drop in recall (-1.98\%), suggesting its suitability for tasks emphasizing prediction reliability.

\textbf{Qwen2:}  
Performed best in recall (2.79\%), reflecting its ability to enhance the coverage of positive samples.

\textbf{GPT:}  
Showed balanced performance with significant gains in accuracy (2.79\%) and precision (2.10\%), making it a versatile choice for multi-domain applications.

\subsection{Experiment Analysis}
\textbf{1. The lower the Basic Filtering Accuracy of a domain, the greater the accuracy improvement from the Self-NLI framework.}

\begin{figure}[!t]
\centering
\includegraphics[width=2.5in]{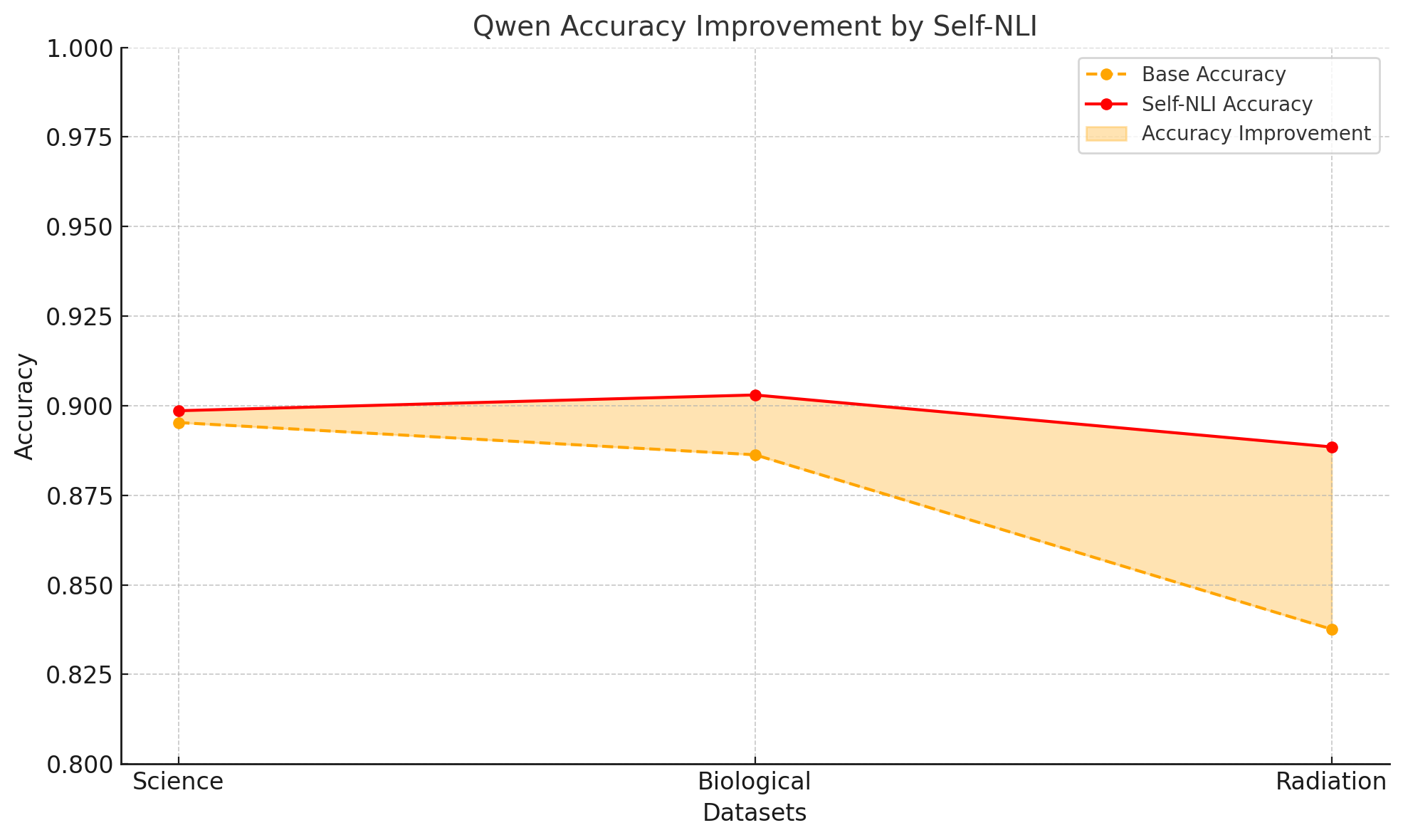}
\captionsetup{justification=raggedright, singlelinecheck=false}
\caption{Qwen Accuracy Improvement}
% \caption{Simulation results for the network.}
\label{fig_5}
\end{figure}

\begin{figure}[!t]
\centering
\includegraphics[width=2.5in]{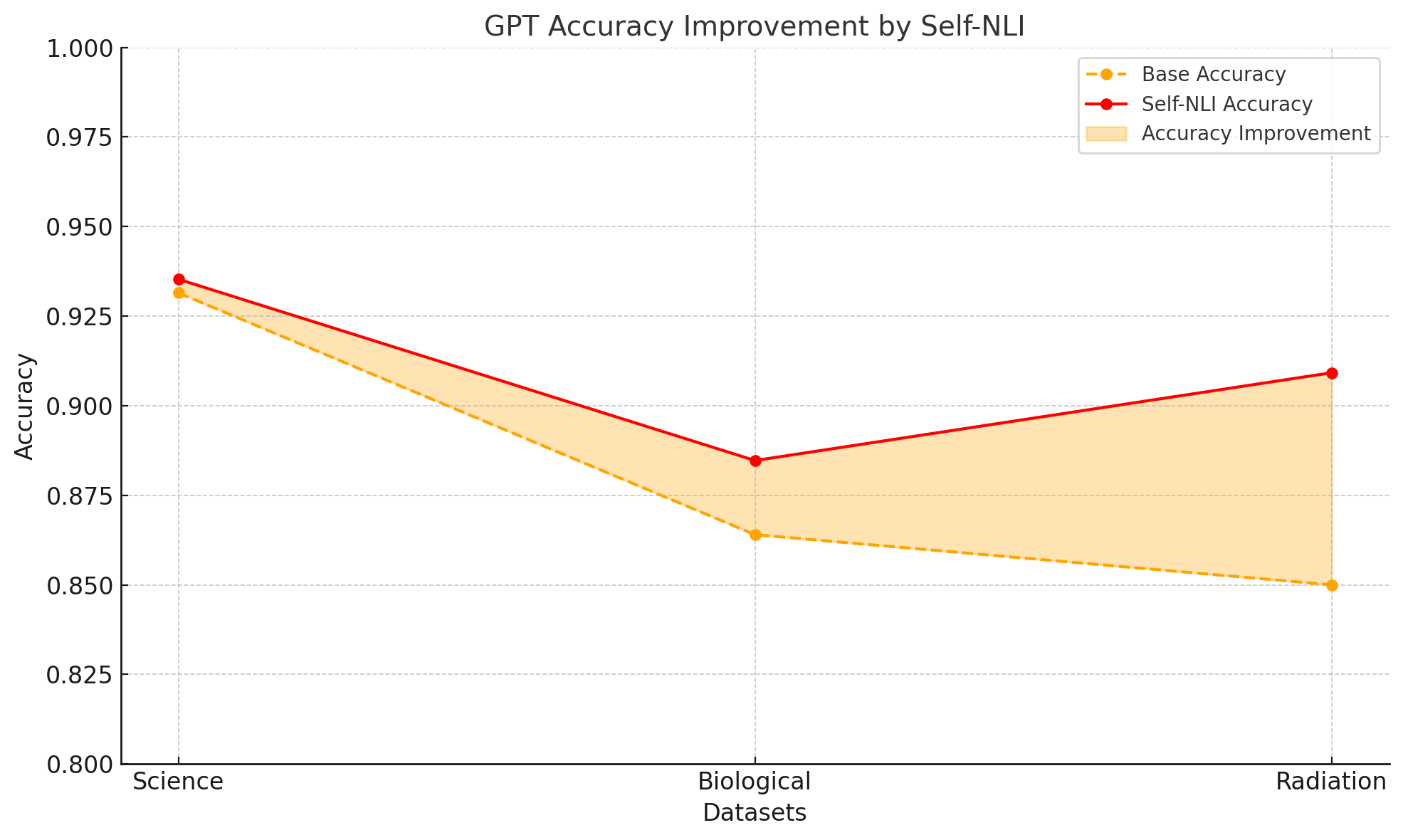}
\captionsetup{justification=raggedright, singlelinecheck=false}
\caption{GPT Accuracy Improvement}
% \caption{Simulation results for the network.}
\label{fig_6}
\end{figure}

Through a comparative analysis of Qwen2 and GPT models across three domain datasets (excluding the RoBERTa model, as it is not a large language model and its prior knowledge was derived from our training)Figure.\ref{fig_5} Figure.\ref{fig_6}, we identified a notable pattern: The lower the Basic Filtering Accuracy of a domain, the more significant the improvement in accuracy achieved by the Self-NLI framework in that domain.  This phenomenon can be explained from the perspective of the prior knowledge of large language models.

Specifically, lower Basic Filtering Accuracy indicates weaker model performance in that domain, reflecting that the model had less exposure to prior data from this domain during training.  Consequently, its understanding and reasoning capabilities in this domain are limited.  In such cases, the Self-NLI framework can compensate for these deficiencies by providing more precise reasoning and filtering mechanisms, thereby significantly enhancing the model's filtering performance.

This finding suggests that the Self-NLI framework demonstrates greater adaptability and compensatory capabilities in domains where large language models have less familiarity.

\textbf{2.Irrelevant or incorrect data significantly reduce filtering performance.}

Comparing the results of the \textit{Fake group} with the \textit{Basic group} reveals that random or incorrect knowledge matching (Fake group) introduces interference in the filtering process. This results in performance degradation or, at best, only minor improvements over the Basic group, while being significantly outperformed by the \textit{Self-NLI group}. For instance:

\textbf{In the Biological dataset}, for the Qwen2 model, the accuracy of the Fake group slightly improves from the Basic group's 0.8863 to 0.8937, but remains lower than the Self-NLI group's 0.9030. For the GPT model, the Fake group accuracy decreases from the Basic group's 0.864 to 0.8323, showing a notable performance drop due to random knowledge matching. Similarly, for the RoBERTa model, the Fake group accuracy drops from 0.8799 in the Basic group to 0.8525, further demonstrating the negative impact of incorrect knowledge matching.

\textbf{In the Science dataset}, the Qwen2 and GPT models show no significant improvement in the Fake group compared to the Basic group (e.g., Fake group accuracy: 0.8900 and 0.9301, Basic group accuracy: 0.8953 and 0.9315). For the RoBERTa model, the F1 score of the Fake group is 0.9359, which is significantly lower than the Self-NLI group's 0.9792.

\textbf{In the Radiation dataset}, the Fake group in the GPT model slightly improves over the Basic group (accuracy: 0.8610 versus 0.85), but it remains well below the Self-NLI group's 0.9092. For the Qwen and RoBERTa models, the Fake group performs worse than the Basic group. For example, in the RoBERTa model, the Fake group accuracy is 0.8588, lower than the Basic group's 0.878.

In summary, the Fake group's results demonstrate that introducing irrelevant or incorrect knowledge matching disrupts the filtering process. Although it may offer slight improvements over the Basic group in certain scenarios, it consistently underperforms compared to accurate knowledge matching in the Self-NLI group. Accurate knowledge matching is critical to achieving superior filtering performance, while incorrect matching can lead to significant degradation.

\textbf{3.Self-NLI filtering reduces the number of clusters, indicating improved data focus} \\
We applied the Self-NLI filtering method and the Basic filtering method to process the data, preprocessed the filtered data using NLTK, and then performed topic clustering with BERTopic.As proposed in the BERTopic paper (Grootendorst, 2022)\cite{ref39}, BERTopic leverages pre-trained Transformer models and clustering algorithms such as HDBSCAN to identify semantically similar groups of documents, facilitating the discovery of cohesive topics.The experimental results are shown in Table \ref{tab:clustering_results}.By comparing the BERTopic clustering results of data processed using the \textit{Self-NLI filtering method} and the \textit{Basic filtering method}, it can be observed that the data filtered by Self-NLI generally result in fewer clusters compared to Basic filtering. This indicates that the Self-NLI filtering method can more effectively remove redundant and irrelevant data, leading to more concentrated data and clearer topic division. This phenomenon is evident across multiple datasets and models, as analyzed in the following.

\textbf{Qwen2 Model:} 
In the \textit{Biological} dataset, the number of clusters produced by Self-NLI (26) is slightly higher than that of Basic (24), showing a minimal difference. However, in the \textit{Science} and \textit{Radiation} datasets, Self-NLI reduces the number of clusters by 1 and 10, respectively, compared to Basic, demonstrating its superior ability to filter redundant topics, especially in the Radiation dataset.

\textbf{GPT-3.5 Model:} 
Self-NLI outperforms Basic filtering across all datasets. In the \textit{Biological} dataset, the number of clusters decreases by 5 (21 versus 16); in the \textit{Science} dataset, it decreases by 4 (7 versus 3); and in the \textit{Radiation} dataset, it decreases by 4 (16 versus 12), which shows better topic compactness.

\textbf{RoBERTa Model:} 
Self-NLI demonstrates cluster optimization across all datasets. The number of clusters in the \textit{Biological}, \textit{Science}, and \textit{Radiation} datasets decreases by 6, 1, and 4, respectively (33 versus. 27, 26 versus. 25, 10 versus. 6), with a significant effect in the Radiation dataset.

In summary, the \textit{Self-NLI filtering method} consistently reduces the number of clusters in BERTopic experiments compared to the \textit{Basic filtering method}, particularly in the Radiation dataset for GPT-3.5 and RoBERTa models and the Science dataset for GPT-3.5. This indicates that the Self-NLI filtering method is more effective in removing irrelevant data, improving data quality, and achieving higher topic compactness in clustering tasks.

\begin{table}[!t]
\caption{Clustering Results After Filtering with Self-NLI and Basic Methods}
\label{tab:clustering_results}
\centering
\renewcommand{\arraystretch}{1.2} % Adjust row height
\setlength{\tabcolsep}{6pt} % Adjust column spacing
\begin{tabular}{lllc}
\toprule
\textbf{Dataset}    & \textbf{Model}  & \textbf{Filtering Method} & \textbf{Number of Clusters} \\ 
\midrule
Biological          & Qwen2           & Self-NLI                  & 26                          \\ 
                    &                 & Basic                     & 24                          \\ 
\midrule
Science             & Qwen2           & Self-NLI                  & 14                          \\ 
                    &                 & Basic                     & 15                          \\ 
\midrule
Radiation           & Qwen2           & Self-NLI                  & 12                          \\ 
                    &                 & Basic                     & 22                          \\ 
\midrule
Biological          & GPT-3.5         & Self-NLI                  & 16                          \\ 
                    &                 & Basic                     & 21                          \\ 
\midrule
Science             & GPT-3.5         & Self-NLI                  & 3                           \\ 
                    &                 & Basic                     & 7                           \\ 
\midrule
Radiation           & GPT-3.5         & Self-NLI                  & 12                          \\ 
                    &                 & Basic                     & 16                          \\ 
\midrule
Biological          & RoBERTa         & Self-NLI                  & 27                          \\ 
                    &                 & Basic                     & 33                          \\ 
\midrule
Science             & RoBERTa         & Self-NLI                  & 26                          \\ 
                    &                 & Basic                     & 26                          \\ 
\midrule
Radiation           & RoBERTa         & Self-NLI                  & 6                           \\ 
                    &                 & Basic                     & 10                          \\ 
\bottomrule
\end{tabular}
\end{table}

\section{CONCLUSION}
\subsection{Limitations and future directions.}
1. The Self-NLI filtering method relies on an initial knowledge base in the vector database. If no domain-specific knowledge is available, manual supplementation of correct data is required to ensure the system's functionality. 2. The approach is inherently dependent on the precision of vector-based knowledge matching. Misaligned or irrelevant matches can undermine filtering effectiveness, as highlighted in Experiment Conclusion 2. 3. The framework's efficacy is closely tied to datasets with intrinsic interconnections. For datasets lacking such relationships, the iterative enhancement of domain understanding fails to operate as intended.
One potential direction for enhancing the Self-NLI framework is to explore alternatives to vector matching for assessing knowledge relevance. While vector matching effectively captures semantic similarity, it may struggle with nuanced or context-specific relationships in domain data. 
The current use of a decision tree model within the Self-NLI framework provides interpretability but may lack the capacity to handle more complex decision-making scenarios.  Replacing the decision tree with more advanced approaches could improve the framework’s performance and scalability.

\subsection{Broader Impact}
The Self-NLI method provides an iterative approach for processing domain-specific data.  It can be used where consistency and iterative optimization are required.  Through continuous filtering, the framework progressively enhances its understanding of the domain, thereby improving filtering performance.  Beyond its application in domain data filtering, this method holds potential for tasks such as text continuation, where iteration and the Self-NLI framework ensure domain specificity and relevance in the generated content.  As demonstrated in the work by Shu et al.\cite{ref40} on Logic-Consistency Text Generation from Semantic Parses, ensuring logical consistency and high-quality generation is critical in text generation tasks.  Their study further proves that recursive approaches can improve the quality of generated text.  This suggests that the Self-NLI framework could potentially be applied in the future to text generation tasks, leveraging its iterative and domain-specific optimization capabilities.

\section{Simple References}
You can manually copy in the resultant .bbl file and set second argument of $\backslash${\tt{begin}} to the number of references
 (used to reserve space for the reference number labels box).

\newpage

% \section{Biography Section}
% If you have an EPS/PDF photo (graphicx package needed), extra braces are
%  needed around the contents of the optional argument to biography to prevent
%  the LaTeX parser from getting confused when it sees the complicated
%  $\backslash${\tt{includegraphics}} command within an optional argument. (You can create
%  your own custom macro containing the $\backslash${\tt{includegraphics}} command to make things
%  simpler here.)
 
% \vspace{11pt}

% \bf{If you include a photo:}\vspace{-33pt}
% \begin{IEEEbiography}[{\includegraphics[width=1in,height=1.25in,clip,keepaspectratio]{fig1}}]{Michael Shell}
% Use $\backslash${\tt{begin\{IEEEbiography\}}} and then for the 1st argument use $\backslash${\tt{includegraphics}} to declare and link the author photo.
% Use the author name as the 3rd argument followed by the biography text.
% \end{IEEEbiography}

% \vspace{11pt}

% \bf{If you will not include a photo:}\vspace{-33pt}
% \begin{IEEEbiographynophoto}{John Doe}
% Use $\backslash${\tt{begin\{IEEEbiographynophoto\}}} and the author name as the argument followed by the biography text.
% \end{IEEEbiographynophoto}

\vfill

\end{document}